\documentstyle[aps,preprint,draft]{revtex}
\title{Solutions of gauge invariant cosmological perturbations in 
 long-wavelength limit}
\author{Yasusada Nambu\footnote{e-mail:~nambu@allegro.phys.nagoya-u.ac.jp} 
and Atsushi Taruya\footnote{e-mail:~ataruya@allegro.phys.nagoya-u.ac.jp}}
\address{Department of Physics, Nagoya University \\ 
        Chikusa, Nagoya 464-01, Japan}
\begin{document}
\maketitle
\begin{abstract}
We investigate gauge invariant cosmological perturbations in a 
spatially flat Friedman-Robertson-Walker  
  universe with scalar fields. It is well known that the evolution equation 
  for the gauge invariant quantities has exact solutions in the 
  long-wavelength limit. We find that these gauge invariant solutions can be 
  obtained by  differentiating the background solution with respect to 
  parameters  contained in the background system. This method is 
  very useful when we analyze the long-wavelength behavior of 
  cosmological perturbation with multiple scalar fields.
\end{abstract}
\vspace{-16.5 cm}
\begin{flushright}
 DPNU-97-48 \\
 November 1997
\end{flushright}
\vspace{16.5 cm}
\def\mpl{m_{pl}}
\def\al{\alpha}
\def\dop#1{\frac{\partial}{\partial#1}}
\def\dfop#1{\frac{\partial F}{\partial#1}}
\def\der#1#2{\frac{\partial#1}{\partial#2}}
\def\tal{\tilde\alpha}
\def\r{{\cal R}}
\def\pa{\phi_{1}}
\def\pb{\phi_{2}}
\def\ca{C_{1}}
\def\cb{C_{2}}
\newpage
\section{introduction}
%
%
%
The theory of gauge invariant cosmological perturbations\cite{bar,ks,mfb} is an important  
tools when we investigate the origin and the evolution of the 
structure in our universe. In the context of general relativity, the 
treatment of linear perturbation of any matter field requires great 
care because these perturbations are in general not gauge invariant. 
We must  evaluate gauge invariant quantities to extract 
physical meaning.

 One of the main application of the gauge invariant perturbation is 
 the theory of inflationary universe, in which the 
gauge invariant treatment of perturbations is crucial. The inflation 
predicts that the seed of density fluctuation, which evolves to the 
structures in our universe, is generated as quantum fluctuation of the 
inflaton field. The created fluctuations is stretched by deSitter 
expansion and their wavelength exceeds the Hubble horizon scale. As 
the gauge dependence of perturbations becomes conspicuous on super 
horizon scale, we must use gauge invariant perturbation to handle 
such a situation.  But the derivation of 
the gauge invariant equation and evaluation of its solution  needs 
tedious calculation in general, especially for 
the model with multiple scalar fields which is considered in the 
context of hybrid or extended inflation and scalar-tensor theories.

We here pay attention to the Einstein gravity with minimally coupled scalar fields 
$\phi$. As a background space, we assume a flat 
Friedman-Robertson-Walker(FRW) universe of which dynamical 
variable is a scale factor $a=e^{\al}$. We use a following gauge 
invariant combination(Mukhanov's variable)\cite{muk}:
\begin{equation}
  Q\equiv\delta\phi-\frac{\dot\phi}{H}\varphi,
\end{equation}
where $\delta\phi$ is a perturbation of the scalar field, 
$\varphi$ is a perturbation of three curvature and $H=\dot\al$ is a Hubble 
parameter. The evolution 
equation for $Q$ is given by
\begin{equation}
  \ddot Q+3H\dot Q+\left[\left(\frac{k}{a}\right)^{2}+
  V_{\phi\phi}-\frac{8\pi G}{e^{3\al}}
  \left(\frac{e^{3\al}}{H}\dot\phi^{2}\right)^{\cdot}\right]Q
  =0.
\end{equation}
It is known that this equation has exact solutions in the 
long-wavelength limit $k\rightarrow 0$\cite{hwa}:
\begin{equation}
   Q=c_{1}\frac{\dot\phi}{H}+c_{2}\frac{\dot\phi}{H}
     \int^{t}\frac{H^{2}}{e^{3\al}\dot\phi^{2}}dt,
\end{equation}
where $c_{1}, c_{2}$ are arbitrary constants. Therefore if we can obtain 
the background solution $(\al(t), \phi(t))$, it is possible to predict the behavior of 
the long-wavelength gauge invariant perturbation without solving the 
evolution equation of the perturbation.

Our question is why we can have the exact solution for $k\rightarrow 0$ 
(0-mode) 
 gauge invariant perturbation. The 0-mode perturbation is a perturbation of 
homogeneous mode, it must be contained in a homogeneous background 
system. So we expect that the evolution equation for the 0-mode 
perturbation can be derived by the analysis of the mini-super space 
model which has no inhomogeneous mode.

In this paper, we aim at clarifying the relation between the gauge invariant 
0-mode solution and the background solution. We will show that the 
evolution equation and the solution of the gauge invariant 0-mode 
perturbation can be obtained within the perturbation of mini-super 
space model. In Sec.II, we use one dimensional autonomous system to 
demonstrate that the perturbed equation and its solution is obtained by 
differentiating the background solution with respect to parameters 
 contained in the background solution. In Sec.III, we treat a 
mini-super space model with  scalar fields and derive the 0-mode 
gauge invariant perturbation equation(Mukhanov's equation) and its 
solutions. We use 
 Hamilton-Jacobi method which is used by several 
 authors\cite{lang,salo} to derive 
 the evolution equation of the 
gauge invariant cosmological perturbation. This method does 
not need to specify any gauge condition during calculation, it is 
suitable to apply to gauge invariant perturbation. Sec.IV is devoted to 
summary. We use the unit in which $c=\hbar=8\pi G=1$ and denote the 
partial derivative of a function $F$ with respect to a some variable 
$A$ by $F_{A}$ throughout the paper.
\section{toy model example}
As a demonstration, we consider a one dimensional autonomous system with a 
Hamiltonian
\begin{equation}
	{\cal H}=\frac{1}{2}p^{2}+V(x).
\end{equation}
Equation of motion is
\begin{equation}
	\ddot x+V'(x)=0.
	\label{ex1}
\end{equation}
The solution of this equation is written as $x=x(t+t_{0})$ where $t_{0}$ 
is an origin of time and an arbitrary constant.
The equation for the perturbation is obtained by splitting 
$x=x^{(0)}+\delta x$ and linearize with respect to $\delta x$:
\begin{equation}
	\ddot{\delta x}+V''(x^{(0)})\delta x=0.
	\label{ex2}
\end{equation}
This equation has the same form as the equation of $\dot x$ which is 
obtained by differentiating the background equation (\ref{ex1}) with respect to 
time variable $t$: $(\dot x)^{\cdot\cdot}+V''(x)\dot x=0$. As $\dot 
x=x_{t_{0}}$, a solution of the perturbation equation is obtained by 
differentiating a background solution with respect to the parameter 
$t_{0}$.

We explain this result more rigorously using  Hamilton-Jacobi(H-J) method 
which is suitable to treat a cosmological model in next section. 
 The H-J equation for a generating function $S$ is
\begin{equation}
	\frac{\partial S}{\partial t}+\frac{1}{2}S_{x}^{2}+V(x)=0,
	\label{HJ1}
\end{equation}
and the evolution equation is
\begin{equation}
	\dot x=S_{x}.
\end{equation}
Eq.(\ref{HJ1}) is solved by $S=-Et+W(x)$($E$ is a separation constant):
\begin{eqnarray}
	&&W_{x}^{2}+2V(x)=2E , \nonumber \\
	&&\dot x=W_{x}.
	\label{HJ2}
\end{eqnarray}
The solution is
\begin{equation}
	t+t_{0}=\int^{x}\frac{dx}{W_{x}}\equiv\tau(x;E),
	\label{S1}
\end{equation}
where $t_{0}$ is an arbitrary constant. From this we get $x=x(t+t_{0};E)$ or 
$E=E(x, t+t_{0})$. The background solution is specified by the two 
parameters $(t_{0},E)$ and the perturbed solution is defined by the 
small change of the constants $(t_{0}+\delta t_{0}, E+\delta E)$.

The generating function is written as $S=S^{(0)}+F$ where $S^{(0)}$ is a 
background part and $F$ is a perturbed part which consists of 
quadratic terms of $\delta t_{0}$ and $\delta E$. $F$ satisfies
\begin{equation}
	\frac{\partial F}{\partial t}+S^{(0)}_{x}F_{x}+\frac{1}{2}F_{x}^{2}=0.
	\label{HJ3}
\end{equation}
Now we change variables from $(x,t)$ to 
$(\tau(x;E),t_{0},E(x,t+t_{0}))$:
\begin{eqnarray}
	\frac{\partial}{\partial t} & = & 
	\frac{\partial\tau}{\partial t}\frac{\partial}{\partial\tau}+\frac{\partial t_{0}}{\partial 
	t}\frac{\partial}{\partial t_{0}}+\frac{\partial E}{\partial 
	t}\frac{\partial}{\partial E}
	\nonumber  \\
	 & = & -\frac{\partial}{\partial t_{0}}+E_{t}\frac{\partial}{\partial 
	 E},
	 \\
	\frac{\partial}{\partial x} & = & \frac{\partial\tau}{\partial 
	x}\frac{\partial}{\partial\tau}+\frac{\partial t_{0}}{\partial 
	x}\frac{\partial}{\partial t_{0}}+\frac{\partial E}{\partial 
	x}\frac{\partial}{\partial E}
	\nonumber \\
	 & = & 
	 \tau_{x}\left(\frac{\partial}{\partial\tau}+\frac{\partial}{\partial 
	 t_{0}}\right)+E_{x}\frac{\partial}{\partial E}.
\end{eqnarray}
The Eq.(\ref{HJ3}) becomes
\begin{equation}
	\frac{\partial F}{\partial\tau}+\frac{1}{2}\left(\tau_{x}\left(\frac{\partial 
	F}{\partial\tau}+\frac{\partial F}{\partial t_{0}}\right)+E_{x}\frac{\partial 
	F}{\partial E}\right)^{2}=0.
\end{equation}
As we are interested in linear perturbation due to the change of the 
parameters $(\delta t_{0},\delta E)$, we replace $\dop{t_{0}}\rightarrow\dop{\delta 
t_{0}}, \dop{E}\rightarrow\dop{\delta E}$.
Using a new variable $\delta t\equiv\delta t_{0}+\tau_{E}\delta 
E$ and taking into account the quadratic form of $F$, the H-J 
equation becomes
\begin{equation}
	\frac{\partial F}{\partial\tau}+\frac{1}{2\dot 
	x^{2}}\left(\frac{\partial F}{\partial\delta t}\right)^{2}=0,
\end{equation}
From this equation, we get a Hamiltonian for the perturbation 
variable $\delta t$:
\begin{equation}
	{\cal H}^{(2)}=\frac{1}{2\dot x^{2}}P_{\delta t}^{2},
\end{equation}
where $P_{\delta t}$ is a conjugate momentum of $\delta t$.
The equation of motion is
\begin{equation}
	\ddot{\delta t}+2\frac{\ddot x}{\dot x}\dot{\delta t}=0.
\end{equation}
The equation for a variable $\delta x\equiv\dot x\delta 
t=x_{t_{0}}\delta t_{0}+x_{E}\delta E$ becomes
\begin{equation}
	\ddot{\delta x}+V''(x)\delta x=0.
\end{equation}
This is our desired result The two independent solutions are given 
by
\begin{equation}
 \delta x=x_{t_{0}},x_{E}.
\end{equation}

The explicit form of the solution is obtained by differentiating 
Eq.(\ref{S1}) with respect $t_{0}$ and $E$. Regarding $x$ as a function 
of $(t+t_{0}, E)$ and differentiating Eq.(\ref{S1}) with respect to 
$t_{0}$ and $E$, 
\begin{eqnarray}
  \dop{t_{0}}:~~1&=&x_{t_{0}}\frac{1}{W_{x}}, \nonumber \\
  \dop{E}:~~0&=&x_{E}\frac{1}{W_{x}}-\int^{x}\frac{W_{xE}}{W_{x}^{2}}dx,
\end{eqnarray}
and we have
\begin{eqnarray}
  x_{t_{0}}&=&W_{x}=\dot x, \nonumber \\
  x_{E}&=&W_{x}\int^{x}\frac{dx}{W_{x}^{3}}=\dot x\int^{t}\frac{dt}{\dot x^{2}}.
\end{eqnarray}
The perturbed solution is obtained by differentiating the background 
solution with respect to the parameters which specify the 
background solution.
\section{gauge invariant long-wavelength solution from mini-super 
space solution}
\subsection{a single scalar field case}

We consider a spatially flat FRW model with a scalar field.  
Dynamical variables are a scale factor $a=e^{\alpha}$ and a scalar 
field $\phi$. The Hamiltonian for this system is
\begin{equation}
  {\cal H}=Ne^{-3\al}\left[-\frac{1}{12}P_{\al}^{2}+
            \frac{1}{2}P_{\phi}^{2}+e^{6\al}V(\phi)\right],
\end{equation}
where $N$ is a lapse function, $P_{\al}$ and $P_{\phi}$ are conjugate 
momentum of $\al$ and $\phi$, $V(\phi)$ is an arbitrary potential for 
the scalar field. The equations of motion are
\begin{eqnarray}
  \dot\al&=&\der{\cal H}{P_{\al}}=-\frac{N}{6}e^{-3\al}P_{\al},~~
  \dot\phi=\der{\cal H}{P_{\phi}}=Ne^{-3\al}P_{\phi}, \nonumber \\
  \dot P_{\al}&=&-\der{\cal H}{\al}=
       Ne^{-3\al}\left[-\frac{1}{4}P_{\al}^{2}+\frac{3}{2}P_{\phi}^{2}-
       3e^{6\al}V(\phi)\right], \\
  \dot P_{\phi}&=&-\der{\cal H}{\phi}=Ne^{3\al}V'(\phi). \nonumber
\end{eqnarray}
The Hamilton-Jacobi equation follows from the Hamiltonian constraint 
with the canonical conjugate momentum replaced by derivative of a 
generating function $S$, $P_{\al}=\der{S}{\al}, P_{\phi}=\der{S}{\phi}$:
\begin{equation}
	-\frac{S_{\al}^{2}}{12}+\frac{S_{\phi}^{2}}{2}+e^{6\al}V(\phi)=0.
\end{equation}
The evolution equations are
\begin{eqnarray}
	\frac{\dot\al}{N} & = & -\frac{e^{-3\al}}{6}S_{\al},
	 \\
	\frac{\dot\phi}{N} & = & e^{-3\al}S_{\phi}.
\end{eqnarray}
The background solution is obtained by assuming 
$S^{(0)}=-2e^{3\al}H(\phi)$. Then
\begin{eqnarray}
	-3H^{2} & + & 2H_{\phi}^{2}+V(\phi)=0,
	\label{HJ4}  \\
	\frac{\dot\al}{N} & = & H,
	\label{ev1}  \\
	\frac{\dot\phi}{N} & = & -2H_{\phi}.
	\label{ev2}
\end{eqnarray}
The solution of Eq.(\ref{HJ4}) has a constant of integration $C$ 
and the solution can be written $H=H(\phi;C)$. By differentiating 
Eq.(\ref{HJ4}) with respect to $C$ and using the evolution equation 
(\ref{ev1}) and (\ref{ev2}), we have
\begin{equation}
  H_{C}=De^{-3\al}\equiv e^{-3(\al+\al_{0})},
\end{equation}
where $D$ and $\al_{0}$ are arbitrary constants. From this, we can 
express $\al+\al_{0}$ as a function of $(\phi, C)$:
\begin{equation}
	\al+\al_{0}\equiv\tilde\al(\phi; C).
\end{equation}
If we invert this 
relation, we get $\phi=\phi(\al+\al_{0}, C)$. $\al_{0}$ is the origin 
of the scale factor and $C$ determines the initial value of the 
scalar field $\phi$. 
 The background solution is 
characterized by two parameters $(\al_{0},C)$ and perturbation is 
defined by small change of these parameters $(\al_{0}+\delta\al_{0},C+\delta C)$.

We split the generating function into the background and the perturbed 
part: $S=S^{(0)}+F$. $F$ satisfies
\begin{equation}	
	-\frac{S^{(0)}_{\al}}{6}F_{\al}+S^{(0)}_{\phi}F_{\phi}-\frac{F_{\al}^{2}}{12}
	+\frac{F_{\phi}^{2}}{2}=0.
	\label{HJ5}
\end{equation}
Change variables from $(\al,\phi)$ to $(\tilde\al,\al_{0},C)$:
\begin{eqnarray}
	\frac{\partial}{\partial\al} & = & 
		\frac{\partial\tilde\al}{\partial\al}\frac{\partial}{\partial\tilde\al}+
	\frac{\partial\al_{0}}{\partial\al}\frac{\partial}{\partial\al_{0}}+
	\frac{\partial C}{\partial\al}\frac{\partial}{\partial C}
	\nonumber \\
	 & = & -\frac{\partial}{\partial\al_{0}}+C_{\al}\frac{\partial}{\partial 
	 C},
	\\
	\frac{\partial}{\partial\phi} & = & 
		\frac{\partial\tilde\al}{\partial\phi}\frac{\partial}{\partial\tilde\al}
	+\frac{\partial\al_{0}}{\partial\phi}\frac{\partial}{\partial\al_{0}}
	+\frac{\partial C}{\partial\phi}\frac{\partial}{\partial C}
	\nonumber \\
	 & = & 
	  \tilde\al_{\phi}\left(\frac{\partial}{\partial\tilde\al}
	  +\frac{\partial}{\partial\al_{0}}\right)+C_{\phi}\frac{\partial}{\partial 
	  C}.
\end{eqnarray}
Then the H-J equation (\ref{HJ5}) becomes
\begin{equation}
	e^{3\al}H\frac{\partial 
	F}{\partial\tilde\al}-\frac{1}{12}\left(-\frac{\partial 
	F}{\partial\al_{0}}+C_{\al}\frac{\partial F}{\partial 
		C}\right)^{2}+\frac{1}{2}
	\left(\tilde\al_{\phi}\left(\frac{\partial F}{\partial\tilde\al}+\frac{\partial 
	F}{\partial\al_{0}}\right)+C_{\phi}\frac{\partial F}{\partial C}\right)^{2}=0.
\end{equation}
To extract the linear perturbation part, we replace 
$\dop{\al_{0}}\rightarrow\dop{\delta\al_{0}}, 
\dop{C}\rightarrow\dop{\delta C}$.
Using a new variable ${\cal 
R}\equiv\delta\alpha_{0}+\tilde\al_{C}\delta C$ and assuming a 
quadratic form of the generating function $F$ with respect to $\r$, we have
\begin{equation}
	\frac{\partial	F}{\partial\tilde\al}
	+\frac{H}{8e^{3\al}H_{\phi}^{2}}\left(\frac{\partial F}{\partial{\cal 
	R}}\right)^{2}=0.
\end{equation}
From this, we get a Hamiltonian for the variable ${\cal R}$:
\begin{equation}
	{\cal H}^{(2)}=\frac{H}{8e^{3\al}H_{\phi}^{2}}P_{\cal R}^{2},
\end{equation}
where $P_{\cal R}$ is a conjugate momentum of ${\cal R}$.
The equation of the motion for ${\cal R}$ becomes
\begin{equation}
	{\cal R}_{\al\al}+\left(3+\frac{H_{\al}}{H}
	+2\frac{\partial}{\partial\al}\ln\left(\frac{H_{\phi}}{H}\right)\right)
	{\cal R}_{\al}=0.
\end{equation}
By introducing a new variable $Q\equiv\phi_{\al}{\cal 
 R}=\phi_{\al_{0}}\delta\al_{0}+\phi_{C}\delta C$, $Q$ 
satisfies
\begin{equation}
	Q_{\al\al}+\left(3+\frac{H_{\al}}{H}\right)Q_{\al}+
	\left(\frac{V_{\phi\phi}}{H^{2}}-\frac{1}{e^{3\al}H}\left(e^{3\al}H\phi_{\al}^{2}\right)_{\al}
	\right)Q=0.
\end{equation}
This is nothing but a Mukhanov equation for the long-wavelength limit. 
For 0-mode perturbation, the only gauge freedom is infinitesimal 
time coordinate transformation $t\rightarrow t+\delta t$. Then 
variables transform as $ \delta\al\rightarrow\delta\al+\dot\al\delta 
t, \delta\phi\rightarrow\delta\phi+\dot\phi\delta t$. A possible gauge 
invariant combination of these variables is 
$\delta\al-(\dot\al/\dot\phi)\delta\phi$ and this has a meaning 
of perturbation of 3 curvature in co-moving gauge (intrinsic 
curvature perturbation). In our 
calculation, we use $\tilde\al$ as a time parameter and this is 
equivalent to using co-moving gauge. Therefore the variable ${\cal R}$ 
is the  gauge invariant intrinsic curvature perturbation or the 
Bardeen's parameter in the long-wavelength limit.

The two independent solutions are given by
\begin{equation}
	Q=\phi_{\al_{0}}, \phi_{C}.
\end{equation}
The explicit form of these solutions can be obtained using 
Eq.(\ref{HJ4}), (\ref{ev1}), (\ref{ev2}). Assuming that the lapse $N$ 
is a function of the scalar field $\phi$, we have
\begin{equation}
  t+t_{0}=-\frac{1}{2}\int^{\phi}\frac{d\phi}{NH_{\phi}}.
                 \label{int}
\end{equation}
Regarding $\phi$ as a function of $(t+t_{0}, C)$ and
differentiating the both side of this equation with respect to $t_{0}$ and $C$,
\begin{eqnarray}
 \dop{t_{0}}:~~1&=&-\frac{\phi_{t_{0}}}{2NH_{\phi}}, \nonumber \\
 \dop{C}:~~0&=&-\frac{\phi_{C}}{2NH_{\phi}}+\int^{\phi}\frac{H_{\phi 
  C}}{2NH_{\phi}^{2}}d\phi.
\end{eqnarray}
Choosing the scale factor as a time parameter($t\equiv\al, N\equiv 1/H$),
\begin{eqnarray}
  \phi_{\al_{0}}&=&-2\frac{H_{\phi}}{H}=\phi_{\al}, \nonumber \\
  \phi_{C}&=&NH_{\phi}\int^{\phi}\frac{H_{\phi C}}{NH_{\phi}^{2}}d\phi=
  -6\phi_{\al}\int^{\al}\frac{e^{-3(\al+\al_{0})}}{\phi_{\al}^{2}H}d\al.
\end{eqnarray}
We therefore obtained the gauge invariant 0-mode solution by 
differentiating the background solution with respect to two 
parameters $\al_{0}$ and $C$.
\subsection{two scalar fields case}
We consider a flat FRW universe with two scalar fields $\pa$ and $\pb$. 
The Hamilton-Jacobi equation is
\begin{equation}
	-\frac{S_{\al}^{2}}{12}+\frac{S_{\pa}^{2}}{2}
	+\frac{S_{\pb}^{2}}{2}+e^{6\al}V(\pa,\pb)=0.
\end{equation}
The background solution is obtained by $S^{(0)}=-2e^{3\al}H(\pa,\pb)$,
\begin{eqnarray}
	-3H^{2} & + & 2H_{\pa}^{2}+2H_{\pb}^{2}+V(\pa,\pb)=0,
	\label{HHJ4}  \\
	\frac{\dot\al}{N} & = & H,
	\label{eev1}  \\
	\frac{\dot\pa}{N} & = & -2H_{\pa},~~\frac{\dot\pb}{N} = -2H_{\pb} .
	\label{eev2}
\end{eqnarray}
The solution of Eq.(\ref{HHJ4}) has two constants of integration 
$C_{1}, C_{2}$ and can be written 
$H=H(\pa,\pb,C_{1},C_{2})$. By differentiating the H-J equation 
(\ref{HHJ4}) with 
respect to these constants and using the evolution equation 
(\ref{eev1}) and (\ref{eev2}), it can be shown
\begin{equation}
	e^{3\al}H_{C1}=D_{1},~~e^{3\al}H_{C2}=D_{2},
\end{equation}
where $D_{1}, D_{2}$ are constants. Let us define new constants $\al_{0}$ 
and  $f$ by 
$D_{1}=e^{-3\al_{0}}, D_{2}=e^{-3\al_{0}}f$. Then
\begin{eqnarray}
	H_{C1} & = & e^{-3(\al+\al_{0})},
	  \\
	\frac{H_{C2}}{H_{C1}}& = & f.
\end{eqnarray}
From this we have
\begin{eqnarray}
	\al+\al_{0} & = & \tilde\al(\pa,\pb,C_{1},C_{2}),
	  \\
	f & = & f(\pa,\pb,C_{1},C_{2}).
\end{eqnarray}
$\al_{0}$ is the origin of scale factor. If $C_{1}, C_{2}$ are fixed, 
 $f$ determines trajectories in configuration space $(\pa, \pb)$.
We split the generating function into the background and the perturbed 
part: $S=S^{(0)}+F$. $F$ satisfies
\begin{equation}	
	-\frac{S^{(0)}_{\al}}{6}F_{\al}+S^{(0)}_{\pa}F_{\pa}+S^{(0)}_{\pb}F_{\pb}-\frac{F_{\al}^{2}}{12}
	+\frac{F_{\pa}^{2}}{2}+\frac{F_{\pb}^{2}}{2}=0.
	\label{HHJ5}
\end{equation}
Change variables from $(\pa,\pb,\al)$ to 
$(\tilde\al,\al_{0},\ca,f,\cb)$:
\begin{eqnarray}
	\dop{\al} & = & -\dop{\al_{0}}+\ca{}_{\al}\dop{\ca}+\cb{}_{\al}\dop{\cb},
	 \\
	\dop{\pa} & = & 
	\tilde\al_{\pa}\left(\dop{\tal}
	+\dop{\al_{0}}\right)+\ca{}_{\pa}\dop{\ca}+f_{\pa}\dop{f}
	+\cb{}_{\pa}\dop{\cb},
	 \\
	\dop{\pb} & = & 
	\tilde\al_{\pb}\left(\dop{\tal}
	+\dop{\al_{0}}\right)+\ca{}_{\pb}\dop{\ca}+f_{\pb}\dop{f}
	+\cb{}_{\pb}\dop{\cb}.
\end{eqnarray}
The H-J equation becomes
\begin{eqnarray}
	e^{3\al}H\dfop{\tal} & - & \frac{1}{12}\left(-\dfop{\al_{0}}+
	\ca{}_{\al}\dfop{\ca}+\cb{}_{\al}\dfop{\cb}\right)^{2}
	\nonumber  \\
	 & + & 
	 \frac{1}{2}\left(\tilde\al_{\pa}\left(\dfop{\tal}
	 +\dfop{\al_{0}}\right)+\ca{}_{\pa}\dfop{\ca}+f_{\pa}\dfop{f}
	 +\cb{}_{\pa}\dfop{\cb}\right)^{2}
      \\
	 & + & 
	 \frac{1}{2}\left(\tilde\al_{\pb}\left(\dfop{\tal}
	 +\dfop{\al_{0}}\right)+\ca{}_{\pb}\dfop{\ca}+f_{\pb}\dfop{f}
	 +\cb{}_{\pb}\dfop{\cb}\right)^{2}=0
	\nonumber
\end{eqnarray}
To extract a linear perturbation part, we replace 
$\dop{\al_{0}}\rightarrow\dop{\delta\al_{0}}, 
\dop{f}\rightarrow\dop{\delta f}, \dop{C_{1}}\rightarrow\dop{\delta 
C_{1}}, \dop{C_{2}}\rightarrow\dop{\delta C_{2}}$. Using a new variable ${\cal 
R}\equiv\delta\al_{0}+\tilde\al_{C1}\delta\ca+\tilde\al_{C2}\delta\cb$ 
and assuming a quadratic form of $F$ with respect to $\r$ and $\delta 
f$, we have
\begin{equation}
	e^{3\al}H\dfop{\tal}
	+\frac{1}{2}\left(\al_{\pa}\dfop{\r}+f_{\pa}\dfop{\delta f}\right)^{2}
	+\frac{1}{2}\left(\al_{\pb}\dfop{\r}+f_{\pb}\dfop{\delta f}\right)^{2}
	=0.
\end{equation}
Define $2\times 2$ matrices
\begin{equation}
\hat X=\left[
\begin{array}{lr}
	\pa{}_{\al}, & \pa{}_{f}  \\
	\pb{}_{\al}, & \pb{}_{f}
\end{array}
\right], ~~
\hat A=\frac{1}{e^{3\al}H}\left(\hat X^{-1}\right)\left(\hat X^{-1}\right)^{T}.
\end{equation}
The H-J equation becomes
\begin{equation}
  \dfop{\tal}+\frac{1}{2}\left[
  \begin{array}{lr}
  \dfop{\r} & \dfop{\delta f} 
  \end{array}\right]
  \hat A \left[
  \begin{array}{c}
    \dfop{\r} \\
    \dfop{\delta f}
  \end{array} \right]=0.
\end{equation}
We get a Hamiltonian for the perturbation variables $(\r, \delta f)$:
\begin{equation}
	{\cal H}^{(2)}=\frac{1}{2}\left[
  \begin{array}{lr}
  P_{\r} & P_{\delta f}
  \end{array}\right]
  \hat A \left[
  \begin{array}{c}
    P_{\r} \\
    P_{\delta f}
  \end{array} \right],
\end{equation}
where $P_{\r}$ and $P_{\delta f}$ are conjugate momentum of $\r$ and 
$\delta f$, respectively.
The evolution equation becomes
\begin{equation}
    \left[
	\begin{array}{c}
	   \r \\ \delta f
	\end{array}
	\right]_{\al\al}
	+\hat A\left(\hat A^{-1}\right)_{\al}
	\left[
	\begin{array}{c}
	   \r \\ \delta f
	\end{array}
	\right]_{\al}=0,
\end{equation}
Introducing a new variable
\begin{equation}
  \vec Q=\left[
    \begin{array}{c}
       Q_{1} \\ Q_{2}
    \end{array}
  \right]
 \equiv 
  \hat X
  \left[
    \begin{array}{c}
       \r \\ \delta f
    \end{array}
  \right],
\end{equation}
$\vec Q$ satisfies
\begin{equation}
  \vec Q_{\al\al}+\left(3+\frac{H_{\al}}{H}\right)\vec Q_{\al}+\hat M 
  \vec Q=0,
\end{equation}
where $\hat M$ is a $2\times 2$ matrix defined by
\begin{equation}
  \left(\hat M\right)_{ij}=
  \frac{V_{\phi_{i}\phi_{j}}}{H^{2}}-
  \frac{1}{e^{3\al}H}
  \left(e^{3\al}H\phi_{i\al}\phi_{j\al}\right)_{\al}.
\end{equation}
This is the Mukhanov equation for two scalar field system in the 
long-wavelength limit\cite{hwang}. The 
four independent solutions of this equation are
\begin{equation}
 \vec Q=\left[
    \begin{array}{c}
       \pa \\ \pb
    \end{array}
  \right]_{\al_{0}},
\left[
    \begin{array}{c}
       \pa \\ \pb
    \end{array}
  \right]_{f},
\left[
    \begin{array}{c}
       \pa \\ \pb
    \end{array}
  \right]_{C_{1}},
\left[
    \begin{array}{c}
       \pa \\ \pb
    \end{array}
  \right]_{C_{2}} \label{qsol}.
\end{equation}
These solutions can be written in another form:
\begin{equation}
 \vec Q=\hat X
 \left[
    \begin{array}{c}
       d_{1} \\ d_{2}
    \end{array}
  \right]
 +\hat X\int^{\al}\frac{1}{e^{3\al}H}\left(\hat X^{-1}\right)
   \left(\hat X^{-1}\right)^{T}d\al
   \left[
    \begin{array}{c}
       d_{3} \\ d_{4}
    \end{array}
  \right], \label{muksol}
\end{equation}
where $d_{1}, d_{2}, d_{3}, d_{4}$ are  arbitrary constants.
Using the Mukhanov's variable, $\r$ and $\delta f$  can be expressed as
\begin{equation}
  \left[
  \begin{array}{c}
    \r \\ \delta f
  \end{array}
  \right]
  =\hat X^{-1}Q=
  \left[
  \begin{array}{c}
     \al_{\pa}Q_{1}+\al_{\pb}Q_{2} \\
     f_{\pa}Q_{1}+f_{\pb}Q_{2}
  \end{array}
  \right].
\end{equation}
In two scalar field case, using $\tilde\al$ as a time parameter means
\begin{equation}
 \delta\tilde\al=\tilde\al_{\pa}\delta\pa+\tilde\al_{\pb}\delta\pb=0.
\end{equation}
From the relation
\begin{equation}
  \tilde\al_{\phi_{i}}=\der{\tilde\al}{H}\der{H}{\phi_{i}}
    =-\frac{H}{2H_{\al}}\phi_{i, \al} \label{alp},
\end{equation}
we have $\delta\tilde\al=\sum_{i}\dot\phi_{i}\delta\phi_{i}=0$. So using $\tilde\al$ as 
a time parameter is equivalent to using co-moving gauge and 
the variable ${\cal R}$ is the gauge invariant intrinsic curvature perturbation in 
the long-wavelength limit.

We can rewrite the expression of ${\cal R}$ using Eq.(\ref{alp}):
\begin{equation}
	{\cal R}=-\frac{H}{2H_{\al}}(\pa{}_{\al}Q_{1}+\pb{}_{\al}Q_{2}).
	\label{bardeen}
\end{equation}
This gives a relation between the curvature perturbation $\r$ 
 which  corresponds to the Bardeen's parameter $\zeta$ and 
 the Mukhanov's variable $Q_{1}, Q_{2}$ in the long-wavelength limit.
 $Q_{1}, Q_{2}$ are given by Eq.(\ref{muksol}) which is written by using 
 background solution. The important point is that we have not assumed 
 any approximation to derive the long-wavelength solution.
 We can get the gauge invariant gravitational potential $\Phi$ by considering 
 perturbation of the evolution equation (\ref{eev1}):
 \begin{equation}
  	\delta\dot\al-H\delta N=\sum_{i}H_{\phi_{i}}\delta\phi_{i}.
  	\label{eev1p}
  \end{equation}
In longitudinal gauge, $\delta N=-\delta\al\equiv\Phi$ and 
$Q_{i}=\delta\phi_{i}+(\dot\phi_{i}/H)\Phi$. The above equation becomes
\begin{equation}
	\dot\Phi+H\Phi=\frac{1}{2}\sum_{i}\dot\phi_{i}\delta\phi_{i}.
	\label{}
\end{equation}
The curvature perturbation ${\cal R}$ 
can be written using $\Phi$:
\begin{eqnarray}
	{\cal R} & = & -\frac{H}{2\dot H}\sum_{i}\dot\phi_{i}Q_{i} \nonumber \\
	 & = & -\frac{H}{2\dot H}\sum_{i}\left(\dot\phi_{i}\delta\phi_{i}+
	 \frac{\dot\phi_{i}^{2}}{H}\Phi\right) \nonumber \\
	 & = & \Phi-\frac{H}{\dot H}(\dot\Phi+H\Phi).
\end{eqnarray}
Therefore we have
\begin{equation}
	\Phi=-\frac{H}{e^{\al}}\int dt e^{\al}\frac{\dot H}{H^{2}}{\cal R}
	\label{newton}
\end{equation}
\subsection{application to inflationary model}
 We demonstrate our 0-mode solution (\ref{qsol}) of perturbation reproduces 
 previously obtained results\cite{po,wands,ms} in inflationary model using the 
 slow rolling approximation. 
 If we concentrate on growing mode solutions, the gauge invariant 
 0-mode solution is given by
\begin{equation}
	\left[
	\begin{array}{l}
	  Q_{1} \\ Q_{2}
	\end{array}
	\right]
	=A
	\left[
	\begin{array}{l}
	  \phi_{1\alpha} \\ \phi_{2\alpha}
	\end{array}
	\right]
	+B
	\left[
	\begin{array}{l}
	  \phi_{1f} \\ \phi_{2f}
	\end{array}
	\right],
\end{equation}
where $A, B$ are arbitrary constants. Substituting this solution to 
 the expression of the curvature perturbation (\ref{bardeen}), we have
\begin{equation}
	{\cal R}=A-B\frac{H}{2H_{\alpha}}\left(
	     \phi_{1\alpha}\phi_{1f}+\phi_{2\alpha}\phi_{2f}\right) 
	     \label{bardb}.
\end{equation}
The first term is the contribution of adiabatic fluctuations and 
remains constant in time. The second term is the contribution due to 
iso-curvature or entropic fluctuations which arises whenever there 
are two or more scalar fields.
Using the slow rolling approximation, H-J equation and evolution 
 equations become
\begin{eqnarray}
	  H^{2}&\approx&\frac{1}{3}V(\phi_{1}, \phi_{2}), \nonumber \\
	  \phi_{1\alpha}&=&-\frac{2}{H}H_{\phi 1}=-\frac{V_{\phi 1}}{V}, 
	  \label{evsl}\\
	  \phi_{2\alpha}&=&-\frac{2}{H}H_{\phi 2}=-\frac{V_{\phi 2}}{V}. 
	  \nonumber 
\end{eqnarray}
 In this approximation, we do not have parameters $C_{1}, C_{2}$ contained 
 in the Hubble function $H$ and we neglect decaying 
 mode solutions of perturbations. In the inflationary phase, $\dot 
 H/H\ll 1$ and the gauge invariant gravitational potential becomes
\begin{equation}
  	\Phi\approx\frac{\dot H}{H^{2}}{\cal R}\approx -\frac{V_{\phi 1}^{2}+
	V_{\phi 2}^{2}}{2V^{2}}{\cal R}.
\end{equation}
For general form of the potential $V(\pa, \pb)$, we can not write down 
the solution of the evolution equation (\ref{evsl}). We consider two examples of 
the potential in which case we have integral form of the slow rolling solution.
	
{\underline{Case 1}}: $V(\phi_{1}, 
\phi_{2})=V_{1}(\phi_{1})+V_{2}(\phi_{2})$. By integrating the evolution equation 
(\ref{evsl}), we have
\begin{eqnarray}
	  \alpha+\alpha_{0}&=&-\int d\phi_{1}\frac{V_{1}}{V_{1, \phi 1}}-\int 
       d\phi_{2}\frac{V_{2}}{V_{2, \phi 2}}, \\
	  f&=&\int\frac{d\phi_{1}}{V_{1,\phi 
	  1}}-\int\frac{d\phi_{2}}{V_{2,\phi 2}},
\end{eqnarray}
 where $\alpha_{0}, f$ are constants of integration. By differentiating the above 
 solution with respect to the parameter $f$, we have
\begin{eqnarray}
	  \phi_{1f}&=&\frac{V_{2}}{V_{1}+V_{2}}V_{1,\phi 1}, \nonumber \\
	  \phi_{2f}&=&-\frac{V_{1}}{V_{1}+V_{2}}V_{2, \phi 2}.
\end{eqnarray}
Substituting these solutions to the expression of ${\cal R}$ 
(\ref{bardb}), we have
\begin{equation}
	{\cal R}=A-B\frac{\left(V_{1, \phi 1}\right)^{2}V_{2}
	           -\left(V_{2, \phi 2}\right)^{2}V_{1}}
	           {\left(V_{1, \phi 1}\right)^{2}+\left(V_{2, \phi 2}\right)^{2}}.
\end{equation}
The gauge invariant gravitational potential is
\begin{equation}
	  \Phi=A\frac{H_{\alpha}}{H}-\frac{B}{2}
	  \frac{V_{1,\alpha}V_{2}-V_{2,\alpha}V_{1}}{V_{1}+V_{2}}.
\end{equation}

{\underline{Case 2}}: $V(\phi_{1}, \phi_{2})=V_{1}(\phi_{1})V_{2}(\phi_{2})$. 
By integrating the evolution equation (\ref{evsl}), we have
\begin{eqnarray}
       \alpha+\alpha_{0}&=&-\int d\phi_{1}\frac{V_{1}}{2V_{1, \phi 1}}-\int 
       d\phi_{2}\frac{V_{2}}{2V_{2, \phi 2}}, \nonumber \\
       f&=&\int d\phi_{1}\frac{V_{1}}{V_{1, \phi 1}}-\int 
       d\phi_{2}\frac{V_{2}}{V_{2, \phi 2}},
\end{eqnarray}
where $\alpha_{0}, f$ are constants of integration. By 
 differentiating the above equations with respect to $f$, we have
\begin{eqnarray}
      \phi_{1, f}&=&\frac{V_{1, \phi 1}}{2V_{1}}, \nonumber \\
      \phi_{2, f}&=&-\frac{V_{2, \phi 2}}{2V_{2}}.
\end{eqnarray}
Substituting these solutions to the expression of ${\cal R}$ 
(\ref{bardb}), we  have
\begin{equation}
     {\cal R}=A-\frac{B}{2}
      \frac{\left(\frac{V_{1, \phi 1}}{V_{1}}\right)^{2}-
             \left(\frac{V_{2, \phi 2}}{V_{2}}\right)^{2}}
             {\left(\frac{V_{1, \phi 1}}{V_{1}}\right)^{2}+
             \left(\frac{V_{2, \phi 2}}{V_{2}}\right)^{2}}.  
\end{equation}	
The gauge invariant gravitational potential is
\begin{equation}
	 \Phi=-\frac{1}{2}\left(A+\frac{B}{2}\right)\left(\frac{V_{1,\phi 
	 1}}{V_{1}}\right)^{2}
	 -\frac{1}{2}\left(A-\frac{B}{2}\right)\left(\frac{V_{2,\phi 
	 2}}{V_{2}}\right)^{2}.
\end{equation}
	
\section{summary}
     We have shown that the solution and the evolution equation of 
    the gauge invariant perturbation in the long-wavelength limit can be 
    derived within the mini-super space model. The solution is 
    obtained by differentiating the background solution with respect 
    to parameters  contained in the background system. These 
    parameters completely specify the background solutions.
     If we need to know  the behavior of the long-wavelength 
    perturbation only, this method is  very useful because we does not have 
    to solve the perturbation equation. What we have to do is to solve 
    the background system and identify the parameters contained in the 
    background solution. The analysis of the 
    long-wavelength perturbation in the model that involves multiple 
    scalar field becomes easier. 
	
	The most important feature of our 0-mode solution is it does not 
	rely on the slow roll approximation which is usually assumed to 
	analyze the perturbation in the multi-field inflationary 
	model\cite{po,wands,ms,sa}. 
	As an application,  a model of reheating after 
    inflation was considered\cite{tn}. The system consists of a oscillating 
    inflaton field  and a massless scalar field which interacts with 
    the inflaton. By using the approximation which can include 
    non-linear effect, we obtained background solution and
     the behavior of the long-wavelength curvature perturbation was 
    obtained using the method presented in this paper.

     As a straightforward extension, it is 
    possible to treat spatially closed and open FRW universe and 
    anisotropic Kasner type universe. 
    Instead of scalar fields, perfect fluid can be included. 
    These subjects  are left as a future exercise.
 
\vspace{1cm}
This work is partly supported by the Grant-In-Aid for Scientific 
Research of the Ministry of Education, Science, Sports and Culture of 
Japan(09740196).

\end{document}